\documentclass{article}

\usepackage{microtype}
\usepackage{graphicx}
\usepackage{subcaption}
\usepackage{booktabs} % for professional tables

\usepackage{hyperref}

\usepackage[accepted]{icml2026}

\usepackage{amsmath}
\usepackage{amssymb}
\usepackage{mathtools}
\usepackage{amsthm}
\usepackage{wasysym}
\usepackage{adjustbox}
\usepackage{amsmath,amsfonts,bm}
\usepackage{relsize}
\usepackage{mathtools}
\usepackage{multirow}

\usepackage[capitalize,noabbrev]{cleveref}

\theoremstyle{plain}

\theoremstyle{definition}

\theoremstyle{remark}

\usepackage[textsize=tiny]{todonotes}

\icmltitlerunning{Physics-Informed Graph Learning Acceleration for Large-Scale AC-OPF with Topology Changes}

\begin{document}

\twocolumn[
  \icmltitle{Physics-Informed Graph Learning Acceleration \\ for Large-Scale AC-OPF with Topology Changes}



  \icmlsetsymbol{equal}{*}

  \begin{icmlauthorlist}
    \icmlauthor{Keunju Song}{yyy}
    \icmlauthor{Kyungnam Park}{yyy}
    \icmlauthor{Sua Choi}{yyy}
    \icmlauthor{Seunguk Kim}{sch}
    \icmlauthor{Tae-un Kim}{sch}
    \icmlauthor{Youngmin Choi}{sch}
    \icmlauthor{Sang-Won Min}{comp}
    \icmlauthor{Hongseok Kim}{yyy}
  \end{icmlauthorlist}

  % \icmlaffiliation{yyy}{Department of Electronic Engineering, Sogang University, Seoul, 04107, Republic of Korea}
  \icmlaffiliation{yyy}{Sogang University, Republic of Korea}
  % \icmlaffiliation{comp}{Korea Electrotechnology Research Institute, Gyeonggi-do, 16029, Republic of Korea}
  \icmlaffiliation{comp}{Korea Electrotechnology Research Institute, Republic of Korea}
  % \icmlaffiliation{sch}{Korea Power Exchange, Jeollanam-do, 58322, Republic of Korea}
  \icmlaffiliation{sch}{Korea Power Exchange, Republic of Korea}

  \icmlcorrespondingauthor{Hongseok Kim}{hongseok@sogang.ac.kr}

  \icmlkeywords{Machine Learning, ICML}

  \vskip 0.3in
]



\printAffiliationsAndNotice{}  % no special notice (required even if empty)

\begin{abstract}
In power systems, alternating current optimal power flow (AC-OPF) has been a challenging problem for decades due to its nonconvexity, but fast and efficient solutions are even more needed because of high penetration of large scale renewable generation and load growth. Recently, neural networks (NN) have gained attention in solving AC-OPF, but it is still in an early stage to be applicable for real and large-scale power system operation with topology-changing characteristics. To end this, we propose a novel framework called GraphOPF that considers topology-adaptability, scalability, NN training time, self-supervision, and feasibility altogether. Extensive experiments show that the proposed framework against the baselines is up to 200 times faster in NN training and up to 66 times faster in solving AC-OPF for large-scale power systems including the real Korean power system, while achieving more than 99\% feasibility.
\end{abstract}

\section{Introduction}
Recently, distributed energy resources (DERs) have been massively deployed in power systems to mitigate climate change while the electric load is rapidly increasing due to the expansion of electric vehicles and data centers. In 2026, DERs are expected to represent 46\% of the world’s electricity generation and the total electricity consumption of data centers will be more than 1,000 TWh \cite{ref_iea}, both of which put power systems under high uncertainty and complex operational conditions. To handle this, fast and reliable operation of the power system becomes crucial, which necessitates solving the full AC optimal power flow (OPF). Ever since 1962 when it was first formulated \cite{ref_opf_1962}, AC-OPF has been a fundamental problem in power system operation, aiming to determine the most cost-effective generator dispatch while satisfying physical constraints. Recently, deep learning (DL)-based approaches have been proposed to solve AC-OPF in a fast and reliable way leveraging new neural network (NN) architectures and learning methods. At an early stage, the works in \cite{ref_deepopf-ac, ref_compact_learning, ref_cnn_opf, ref_trans_cnn_opf} utilized modern NN architectures, e.g., deep neural networks (DNN), convolutional neural networks (CNN), and Transformers. Very recently, graph neural networks (GNN) have been studied to solve more practical AC-OPF problems (e.g., security-constrained AC-OPF and AC-OPF with topology changes). In \cite{ref_gnn_opf1}, the authors showed that GNN is effective for topology changes compared to DNN and CNN models. Then, the works in \cite{ref_gnn_opf2} and \cite{ref_gnn_opf6} analyzed the properties of GNN and proposed a GNN-based learning method for topology changes while keeping model complexity low. Meanwhile, the authors of \cite{ref_gnn_opf4} focused on the model architecture, designing the physics-embedded GNN that follows the Gauss-Seidel iteration of AC power flow to solve AC-OPF in \(N-1\) contingency.

Other than neural network architectures, learning mechanism is another considerable factor in solving AC-OPF. To minimize the power generation cost and satisfy the physical constraints simultaneously, the physics-informed learning has been studied in two categories: supervised learning and unsupervised learning. The works in \cite{ref_deepopf-V, ref_compact_learning, ref_cnn_opf, ref_trans_cnn_opf, ref_gnn_opf1} designed the loss function as mean squared error or mean absolute error, which require labels. However, they may incur expensive costs for NN training because they need labels generated from conventional solvers. Semi-supervised learning \cite{hienalternative, ref_opf_pinn_acc} has been considered to mitigate this, however, it cannot be fully resolved. To overcome this, the works in \cite{ref_dc3, ref_deeplde} designed an unsupervised learning method by leveraging the implicit function theorem to enable NN training without labels while satisfying the constraints.

Even though the aforementioned works showed the potentials of NNs in solving AC-OPF, five practical perspectives such as training time, scalability, feasibility, self supervision, and topology adaptability were not considered altogether because of their trade-offs. In order to overcome all the challenges, we propose GraphOPF, a topology-adaptable unsupervised GNN framework that enables scalable and fast learning. Specifically, GraphOPF utilizes edge-aided GNN (EA-GNN) and a hard-constrained embedded layer. Unlike conventional GNN architecture, EA-GNN consists of multi-edge convolutional layer and Chebyshev graph convolutional layer, which consider edge
features of graph data to improve the solution performance (e.g., optimality gap, feasibility) and convergence of NN training further. Moreover, to make it compatible for large-scale AC-OPF, the hard-constrained embedded layer is integrated with EA-GNN to overcome the heavy computational
burden of computing AC power flow in implicit layer. Along with those functionalities, the proposed GraphOPF follows an unsupervised learning scheme, so it does not require extra time in preparing labels.

\begin{figure*}
\centering
\includegraphics[width=\textwidth]{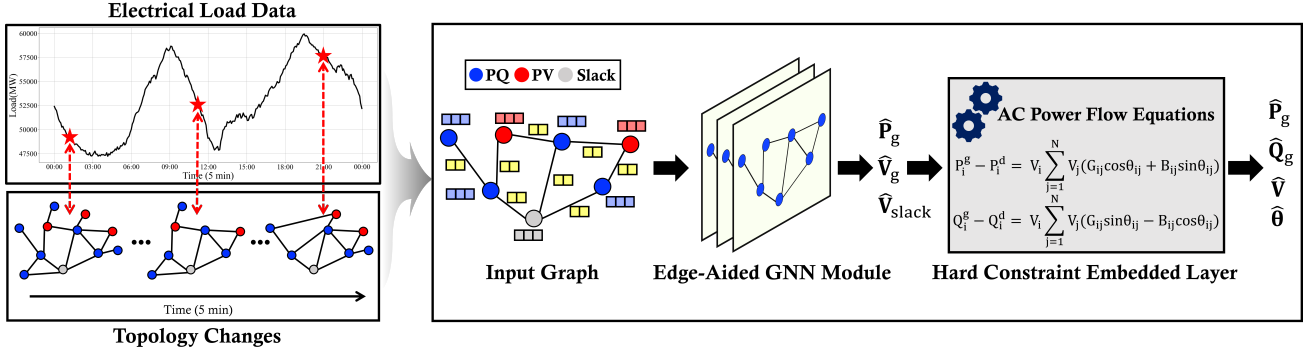}
\caption{The overall framework of the proposed GraphOPF.}
\label{graphopf-framework}
\end{figure*}

We summarize our key contributions as follows.
\begin{itemize}
\item {We propose a novel GNN architecture called EA-GNN, which effectively solves AC-OPF problems. By leveraging edge features along with node features of power systems, EA-GNN maintains the quality of AC-OPF solutions (i.e., optimality gap and feasibility) even for national scale power systems. Furthermore, thanks to the graph representation learning in EA-GNN, the proposed GraphOPF is adaptable for topology changes.}

\item {To make it applicable for large-scale AC-OPF, we design a hard-constrained embedded layer to accelerate computation for power flow equations. This ensures that equality constraints in AC-OPF are strictly satisfied. We find that integrating EA-GNN with a hard-constrained embedded layer enables fast convergence in NN training without using labels.}

\item{The proposed GraphOPF outperforms baselines in inference time, NN training time, and constraint satisfaction for large-scale AC-OPF. Specifically, for the real Korean power system having around 4500 buses and 6000 transmission lines, our framework takes around \textit{five minutes} for NN training and shows 99\% or more constraint satisfaction on average. Furthermore, even under topology changes during real power system operation, GraphOPF can finish model update within \textit{one minute} by using graph transfer learning.}
\end{itemize}

The rest of this paper is organized as follows. In Section~\ref{section:ac-opf}, we describe the AC-OPF problem formulation. In Section~\ref{section:proposed_method}, we present the methodologies of GraphOPF. The experimental settings and results are given in Section~\ref{section:experimental_settings} and Section~\ref{section:experimental_results}, followed by the conclusion in Section~\ref{section:conclusion}.

\section{Problem Formulation} \label{section:ac-opf}
The AC-OPF problem, a nonlinear and nonconvex optimization problem in power systems, aims to minimize the power generation cost while satisfying the physical constraints. For a power network \(\mathcal{G} = (\mathcal{N}, \mathcal{E})\) where \(\mathcal{N}\) is the set of buses and \(\mathcal{E}\) is the set of transmission lines, AC-OPF is formulated as follows:
\begin{subequations}\label{eq:AC-OPF}
\begin{align}
& \min_{\substack{\mathbf{P}_g, \mathbf{Q}_g}, \mathbf{V}, \bm{\theta}}\quad \sum_{i\in \mathcal{N}} C_i(P_{g,i})\label{eq:generation_cost}\\ 
& \text{s.t.} \enspace P_{g,i}^{\min}\leq P_{g,i}\leq P_{g,i}^{\max}, \forall i\in\mathcal{N},\label{eq:generator_active_limit}\\
% &\text{}
& \enspace\quad Q_{g,i}^{\min}\leq Q_{g,i}\leq Q_{g,i}^{\max}, \forall i\in\mathcal{N},\label{eq:generator_reactive_limit}\\
% &\text{}
& \enspace\quad V_i^{\min}\leq V_i \leq V_i^{\max}, \forall i\in\mathcal{N}, \label{eq:voltage_magnitude_limit}\\
% &\text{}
& \enspace\quad p_{ij}^2 + q_{ij}^2 \leq (s_{ij}^{\max})^2, \forall(i,j)\in \mathcal{E}\label{eq:branch_flow_limit},\\
% &\text{}
& \enspace\quad p_{ij} = -G_{ij}V_{i}^2+V_iV_j(G_{ij}\cos\theta_{ij}+B_{ij}\sin\theta_{ij})\label{eq:active_branch_power_flow},\\
& \enspace\quad q_{ij} = B_{ij}V_{i}^2+V_iV_j(G_{ij}\sin\theta_{ij}-B_{ij}\cos\theta_{ij})\label{eq:reactive_branch_power_flow}, \\
& \enspace\quad P_{g,i} - P_{d,i} = V_i\sum_{j=1}^{N}V_j(G_{ij}\cos\theta_{ij}+B_{ij}\sin\theta_{ij})\label{eq:active_power_flow}, \\
& \enspace\quad Q_{g,i} - Q_{d,i} = V_i\sum_{j=1}^{N}V_j(G_{ij}\sin\theta_{ij}-B_{ij}\cos\theta_{ij})\label{eq:reactive_power_flow},
\end{align}
\end{subequations}

where the optimization variables \(\mathbf{P}_g, \mathbf{Q}_{g}, \mathbf{V}, \hspace{1mm} \bm{\theta}\) are in \(\mathbb{R}^{|\mathcal{N}|}\) where \(|\cdot|\) is the cardinality of a set, and their \(i\)-th element is associated with bus \(i\in\mathcal{N}\). For example, \(P_{g,i}\) and \(Q_{g,i}\) are the active and reactive power of generator \(i \in \mathcal{N}\), \(P_{d,i}\) and \(Q_{d,i}\) are the active and reactive load demands of bus \(i \in \mathcal{N}\), \(V_i\) is the voltage magnitude of bus \(i \in \mathcal{N}\), and \(\theta_{ij} = \theta_i - \theta_j\) is the voltage angle differences between bus \(i\) and bus \(j\). Let \(p_{ij}\) and \(q_{ij}\) denote the active and reactive branch power flows from bus \(i\) to bus \(j\). Let \(G_{ij}\) and \(B_{ij}\) denote the conductance and susceptance of the nodal admittance matrix \(\mathbf{Y} \in \mathbb{R}^{|\mathcal{N}| \times |\mathcal{N}|} \). The objective function (\ref{eq:generation_cost}) is the sum of generator cost functions \(C_i(P_{g,i}), i \in \mathcal{N}\). The constraints (\ref{eq:generator_active_limit}) and (\ref{eq:generator_reactive_limit}) limit the generator output of active and reactive power, (\ref{eq:voltage_magnitude_limit}) enforces that voltage magnitudes remain within their specified limits, (\ref{eq:branch_flow_limit}) is the line flow limit, (\ref{eq:active_branch_power_flow}) and (\ref{eq:reactive_branch_power_flow}) indicate the branch power equations based on Ohm's law, and the constraints (\ref{eq:active_power_flow}) and (\ref{eq:reactive_power_flow}) are power flow equations based on Kirchhoff's current law.

\section{Proposed Methodologies} \label{section:proposed_method}
In this section, we describe the overall process of GraphOPF as shown in \cref{graphopf-framework}.

\subsection{Edge-Aided Graph Neural Networks} \label{subsection:EAGNN}
The input graph of GNN is a power network \(\mathcal{G} = (\mathcal{N}, \mathcal{E})\) where the nodes and edges correspond to the buses and transmission lines. Then, a node feature matrix \(\mathbf{X}^{(0)}\in \mathbb{R}^{|\mathcal{N}| \times 2}\) consists of the active and reactive load demands of bus \(i \in \mathcal{N}\). In this study, we leverage an edge feature vector \(\mathbf{E}\in \mathbb{R}^{|\mathcal{E}| \times 4}\) that consists of the resistance, reactance, capacitance, and the line thermal limit of transmission lines. To handle the node and edge features collectively, we propose EA-GNN by leveraging multi-edge convolutional layer and Chebyshev graph convolutional layer as shown in Fig~\ref{graphopf-eagnn}. Details are as follows.

\subsubsection{Multi-Edge Convolutional Layer}
A multi-edge convolutional layer (MEConv) is the first layer that has the graph data \(\mathbf{X}^{(0)}=\{\mathbf{x}_{i}^{(0)}\mid i\in\mathcal{N}\}\)~and \(\mathbf{E}\) as inputs. This layer follows the message passing framework~\cite{ref_gnn_book}, which is the common mechanism in GNN. Then, the proposed MEConv layer is given by

\begin{subequations}\label{meconv_layer}
    \begin{align}
       &\mu_{ij}=\mathlarger{\mathlarger{\Phi}}\big(\kappa\mathbf{x}_{i}^{(l-1)}||\kappa\mathbf{x}_{j}^{(l-1)}||\kappa e_{(i,j)};\Theta_{\text{edge}}^{(l)}\big), \label{agg_func} \\
       &\mu_{i} = {1\over{|\mathcal{N}_i}|}\sum_{j\in\mathcal{N}_{i}}\mu_{ij}, \label{mean_aggr} \\
       & \sigma_{i} = \sqrt{{1\over{|\mathcal{N}_i}|}\sum_{j\in\mathcal{N}_{i}}\mu_{ij}^{2} -\mu_{i}^{2}}, \label{std_aggr} \\
       &\mathbf{x}_{i}^{(l)} = \prod\big(\mu_{i}+\sigma_{i};\Theta_{\text{aggr}}^{(l)}\big), \label{meconv_layer_}
    \end{align}
\end{subequations}

\noindent where \(\mathlarger{\mathlarger{\Phi}}\) is the aggregate function parameterized by \(\Theta_{\text{edge}}^{(l)}\) having two multi-layer perceptrons (MLPs), \(\kappa = {1\over\sqrt{\text{deg}(i)\text{deg(j)}}}\) denotes the normalization coefficient for the degree of node \(i\) and \(j\), and \(||\) denotes concatenation. For the message function, we propose a multi-moment aggregation method by designing the projection layer \(\prod\) parameterized by \(\Theta_{\text{aggr}}^{(l)}\), which takes the sum of mean and standard deviation of the outputs of \(\mathlarger{\mathlarger{\Phi}}\) as an input. Surprisingly, we will see that having multi-moments, i.e., the first and second moments of the distribution of the aggregate function \(\mathlarger{\mathlarger{\Phi}}\) generates, is simple but effective.

Through a multi-moment aggregation method with NNs, our proposed framework first outputs the \textit{topological} feature matrix of power systems for every node \(\mathbf{X}_{\text{MEC}} \in \mathbb{R}^{|\mathcal{N}|\times F}\). Note that \(\mathbf{X}_{\text{MEC}}\) is a \textit{node} feature matrix derived from the node feature matrix \(\mathbf{X}^{(l)}\) and the edge feature vector \(\mathbf{E}\), and can be described as the assembled MEConv in the graph perspective such as

\begin{equation} \label{meconv}
    \mathbf{X}^{(l)}_{\text{MEC}} = MEConv\big( \mathbf{X}^{(l)}, \mathbf{E} ; \Theta_{\text{edge}}^{(l)}, \Theta_{\text{aggr}}^{(l)} \big).
\end{equation}

\subsubsection{Chebyshev Graph Convolutional Layer} After getting the topological feature \(\mathbf{X}^{(l)}_{\text{MEC}}\) from MEConv, we feed it as input to the Chebyshev graph convolutional layer (ChebyConv)~\cite{ref_chebconv_gnn} which is parameterized by \(\Theta^{(l)}_{\text{Cheby}}=\{\Theta_k^{(l)}\mid k=1,\cdots,K\}\) so that

\begin{equation} \label{chebyconv}
        \mathbf{X}^{(l)}_{\text{Cheby}} =  ChebyConv(\mathbf{X}^{(l)}_{\text{MEC}};\Theta^{(l)}_{\text{Cheby}})
\end{equation}

\noindent where \(\mathbf{X}^{(l)}_{\text{Cheby}}\in\mathbb{R}^{|\mathcal{N}|\times F}\) is the output node feature matrix of the ChebConv layer. Note that \(\mathbf{X}^{(l)}_{\text{MEC}}\) from MEConv only considers 1-hop neighbors of each node, so it relies on local connectivity features. Therefore, through the ChebyConv layer, we consider \(K\)-hop neighbors (i.e., \(K\) Chebyshev filters) of each node for extracting global connectivity features. A ChebyConv layer follows the Chebyshev polynomials,

\begin{figure*}[t]
\centering
\includegraphics[width=\textwidth]{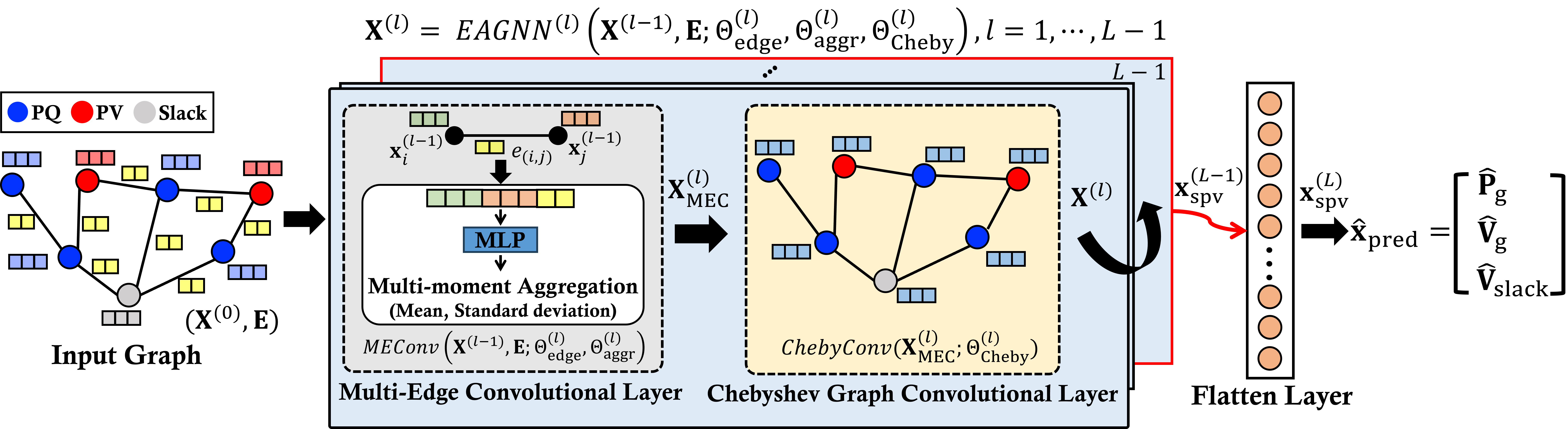}
\caption{The architecture of the proposed EA-GNN.}
\label{graphopf-eagnn}
\end{figure*}

\begin{equation}
    \mathbf{X}^{(l)}_{\text{Cheby}} = \sum_{k=1}^{K}\mathbf{Z}_{k}^{(l)}\Theta_{k}^{(l)}
\end{equation}

\noindent where \(K\) corresponds to the Chebyshev filter size, and \(\mathbf{Z}_{k}^{(l)}\) is computed recursively as \(\mathbf{Z}_{k}^{(l)}=2\hat{\mathbf{L}}\mathbf{Z}_{k-1}^{(l)}-\mathbf{Z}_{k-2}^{(l)}\) where \(\mathbf{Z}_{0}^{(l)} = \mathbf{X}_{\text{MEC}}^{(l)}\) and \(\mathbf{Z}_{1}^{(l)} = \hat{\mathbf{L}}\mathbf{X}^{(l)}_{\text{MEC}}\). \(\hat{\mathbf{L}}={2\mathbf{L}\over{\lambda_{\text{max}}}}-\mathbf{I}\) is the scaled and normalized Laplacian where \(\mathbf{L} = \mathbf{D}-\mathbf{A}\) is the Laplacian matrix, which is calculated by the diagonal degree matrix \(\mathbf{D} \in \mathbb{R}^{|\mathcal{N}|\times|\mathcal{N}|}\) and the adjacency matrix \(\mathbf{A} \in \mathbb{R}^{|\mathcal{N}|\times|\mathcal{N}|}\). \(\lambda_{\text{max}}\) denotes the largest eigenvalue of \(\mathbf{L}\) and \(\mathbf{I} \in \mathbb{R}^{|\mathcal{N}|\times|\mathcal{N}|} \) denotes the identity matrix.

By combining the MEConv and ChebyConv layers sequentially, now we can define the assembled EA-GNN that effectively captures the \textit{interactive} topological features \(\mathbf{X}^{(l)}\) considering local and global connectivity in the graph data:
\begin{equation} \label{eagnn_def}
    \begin{split}
        \mathbf{X}^{(l)} &= EAGNN^{(l)}(\mathbf{X}^{(l-1)},\mathbf{E};\Theta_{\text{edge}}^{(l)},\Theta_{\text{aggr}}^{(l)},\Theta_{\text{Cheby}}^{(l)}),\\&l=1,\dots,L-1.
    \end{split}
\end{equation}
\subsubsection{Flatten Layer}
For the topological features of EA-GNN at the last layer \(\mathbf{X}^{(L-1)}\in\mathbb{R}^{|\mathcal{N}|\times F_{\text{out}}}\), we use the reshape function \(f_{\text{reshape}}:\mathbb{R}^{|\mathcal{N}|\times F_{\text{out}}}\rightarrow{\mathbb{R}^{F_{\text{spv}}}}\) to extract the features of generator and slack bus in a power network, where \(F_{\text{spv}} = |\mathcal{N}_{\text{spv}}|\times F_{\text{out}}\) and \(\mathcal{N}_{\text{spv}} \subseteq \mathcal{N}\) is a union of a set of generator buses \(\mathcal{N}_{\text{pv}}\) and slack bus \(\mathcal{N}_{\text{slack}}\):

\begin{equation} \label{reshape_func}
    \mathbf{x}_{\text{spv}}^{(L-1)} = f_{\text{reshape}}(\mathbf{X}^{(L-1)})
\end{equation}

\noindent where \(\mathbf{x}_{\text{spv}}^{(L-1)}\) is the partial topological features of \(\mathbf{X}^{(L-1)}\) that consist of generator and slack bus. Then, we input \(\mathbf{x}_{\text{spv}}^{(L-1)}\) into the flatten layer to output \(\mathbf{x}_{\text{spv}}^{(L)} \in \mathbb{R}^{2 |\mathcal{N}_{\text{pv}}| + |\mathcal{N}_{\text{slack}}|}\) so that

\begin{equation} \label{flatten_layer}
     \mathbf{x}_{\text{spv}}^{(L)} = Flatten(\mathbf{x}_{\text{spv}}^{(L-1)};\Theta_{\text{flat}}).   
\end{equation}

In (\ref{flatten_layer}), \(\mathbf{x}_{\text{spv}}^{(L)}\) is within the range of 0 and 1 because a sigmoid function is used as an activation function. Thus, we can finally reconstruct \(\hat{\mathbf{x}}_{\text{pred}}\) from \(\mathbf{x}_{\text{spv}}^{(L)}\) by using operating constraints \(\mathbf{x}^{\max}, \mathbf{x}^{\min} \in \mathbb{R}^{2 |\mathcal{N}_{\text{pv}}| + |\mathcal{N}_{\text{slack}}|}\) (i.e., the limit of active power and voltage magnitude of generator and the slack bus), which is the \textit{partial} decision variables in (\ref{eq:AC-OPF}):

\begin{equation} \label{recon_output}
    \hat{\mathbf{x}}_{\text{pred}} = \mathbf{x}_{\text{spv}}^{(L)}(\mathbf{x}^{\max}) + (\mathbf{1}-\mathbf{x}_{\text{spv}}^{(L)})(\mathbf{x}^{\min}).
\end{equation}

Note that \(\hat{\mathbf{x}}_{\text{pred}} = [\hat{\mathbf{P}}_{g},\hat{\mathbf{V}}_{g},\hat{\mathbf{V}}_{\text{slack}}]^{\intercal}\) consists of the active power generation and voltage magnitude for generator bus and voltage magnitude for slack bus, respectively.

\subsection{Hard Constraint Embedded Layer} % Power Flow Equations, implicit layer
So far, we have discussed the proposed EA-GNN module as summarized in Fig~\ref{graphopf-eagnn}. Now we discuss how to output \textit{full} decision variables in (\ref{eq:AC-OPF}) based on the \textit{partial} variables from EA-GNN by using the proposed hard constraint embedded layer, as shown in Fig~\ref{graphopf-framework}. Specifically, we exploit the advanced implicit layer that can accelerate the computation, having \(\hat{\mathbf{x}}_{\text{pred}}\) in (\ref{recon_output}) as input.

\subsubsection{Advanced Implicit Layer}
First, to formulate the hard constraint embedded layer, we express the AC-OPF problem (\ref{eq:AC-OPF}) into the standard constrained optimization problem:

\begin{equation}
    \min_{\mathbf{y}\in\mathbb{R}^{n}} f_{\mathbf{d}}(\mathbf{y}) \quad \text{s.t.} \; \space g_{\mathbf{d}}(\mathbf{y})\leq0,\;  h_{\mathbf{d}}(\mathbf{y})=0
\end{equation}

\noindent where \(\mathbf{d}\in\mathbb{R}^{n} \sim \mathcal{D}\) is the input data point from the dataset \(\mathcal{D}\), \(\mathbf{y} \in \mathbb{R}^{n}\) has the optimal decision variables, \(f_{\mathbf{d}}:\mathbb{R}^{n}\rightarrow{\mathbb{R}}\) is the objective function, \(g_{\mathbf{d}}:\mathbb{R}^{n}\rightarrow{\mathbb{R}^{n_{\text{ineq}}}}\) denotes the inequality constraints, and \(h_{\mathbf{d}}:\mathbb{R}^{n}\rightarrow{\mathbb{R}^{n_{\text{eq}}}}\) denotes the equality constraints.

In our problem, \(\mathbf{y} = [\mathbf{P}_{g},\mathbf{Q}_{g},\mathbf{V},\bm{\theta}]^{\intercal}\) is a collection of the decision variables in (\ref{eq:AC-OPF}), \(g_{\mathbf{d}}\) corresponds to (\ref{eq:generator_active_limit})$-$(\ref{eq:branch_flow_limit}), and \(h_{\mathbf{d}}\) corresponds to (\ref{eq:active_power_flow}) and (\ref{eq:reactive_power_flow}), which are nonlinear and make our problem nonconvex optimization~\cite{ref_ac_np_hard}. Thus, satisfying \(h_{\mathbf{d}}\) by using NN is a challenging part as mentioned in \cite{ref_deepopf-V, ref_eq_hard_nn_opf}. To overcome this problem, the implicit layer method has been considered to ensure that the equality constraints of \(h_{\mathbf{d}}\) are always satisfied \cite{ref_dc3, ref_deeplde, ref_opf_pinn_acc}. In our work, the proposed hard constraint embedded layer also follows the implicit layer method. However, directly calculating the Jacobian inverse in the Newton updates of the implicit layer \cite{ref_dc3, ref_deeplde} becomes computationally expensive for large-scale problems. To address this issue, we leverage the associated linear systems using LU factorization with partial pivoting. We find that this is simple but effectively accelerates the computation of \(h_{\mathbf{d}}\) of a large-scale power system while considering implicit differentiation \cite{ref_dc3, ref_deeplde}, resulting in an additional reduction of the overall NN training time.

\subsection{Lagrangian Dual-based Loss Function} \label{section:LD_loss}
Due to the hard constraint embedded layer, we can define a Lagrangian dual-based loss function~\cite{ref_deeplde} without considering equality constraints as follows:
\begin{equation} \label{loss_func}
    \mathcal{L}_{\text{loss}}(\hat{\mathbf{y}}) = \mathcal{L}_{\text{obj}}(\hat{\mathbf{y}}) + \mathcal{L}_{\text{ineq}}(\hat{\mathbf{y}})
\end{equation}

\noindent where \(\hat{\mathbf{y}} = [\hat{\mathbf{P}}_{g},\hat{\mathbf{Q}}_{g},\hat{\mathbf{V}},\hat{\bm{\theta}}]^{\intercal}\), \(\mathcal{L}_{\text{obj}}\) is the cost of generation by the predicted active power generation \(\hat{\mathbf{P}}_{g}\), and \(\mathcal{L}_{\text{ineq}}\) denotes the amount of total violation. Specifically, with the Lagrange multipliers for each instance in a set of inequality constraints \(\mathcal{C}_{\text{ineq}}\), we have \(\mathcal{L}_{\text{ineq}}(\hat{\mathbf{y}}) = \sum_{c\in\mathcal{C}_{\text{ineq}}}\lambda_{c}\max(0,\nu_{c}(\hat{\mathbf{y}}))\) where \(\nu_c({\hat{\mathbf{y}}})\) can be determined as
\begin{subequations}\label{vio_degrees}
    \begin{align}
       &\nu_{c}^{L}(\hat{P}_{g,i}) = (P_{g,i}^{\min}-\hat{P}_{g,i}), \enspace \forall i\in \mathcal{N}, \label{vio_degree_pg_low} \\
       &\nu_{c}^{U}(\hat{P}_{g,i}) = (\hat{P}_{g,i}-P_{g,i}^{\max}), \enspace \forall i\in \mathcal{N}, \label{vio_degree_pg_up} \\
       &\nu_{c}^{L}(\hat{Q}_{g,i}) = (Q_{g,i}^{\min}-\hat{Q}_{g,i}), \enspace \forall i\in \mathcal{N}, \label{vio_degree_qg_low} \\
       & \nu_{c}^{U}(\hat{Q}_{g,i}) = (\hat{Q}_{g,i}-Q_{g,i}^{\max}), \enspace \forall i\in \mathcal{N}, \label{vio_degree_qg_up} \\
       &\nu_{c}^{L}(\hat{V_{i}}) = (V_{i}^{\min}-\hat{V}_{i}), \enspace \forall i\in \mathcal{N}, \label{vio_degree_v_low} \\ 
       &\nu_{c}^{U}(\hat{V}_{i}) = (\hat{V}_{i}-V^{\max}_{i}), \enspace \forall i\in \mathcal{N}, \label{vio_degree_v_up} \\
       &\nu_{c}^{L}(\hat{s}_{(i,j)}) = (\hat{s}_{(i,j)}-s^{\max}_{(i,j)}), \enspace \forall (i,j)\in \mathcal{E}, \label{vio_degree_s} 
    \end{align}
\end{subequations}
\noindent where \(\hat{\mathbf{s}}_{(i,j)}\) is the predicted branch power flow over transmission lines, which can be computed by (\ref{eq:active_branch_power_flow})$-$(\ref{eq:reactive_branch_power_flow}) based on \(\hat{V_{i}}\) and \(\hat{\theta}_{ij}\). In doing this, (\ref{vio_degree_pg_low})$-$(\ref{vio_degree_s}) induce to satisfy the inequality constraints in AC-OPF (\ref{eq:generator_active_limit})$-$(\ref{eq:branch_flow_limit}) during NN training. Finally, we update \(\Theta_{\text{edge}}, \Theta_{\text{aggr}}, \Theta_{\text{Cheby}}, \Theta_{\text{flat}}\) (i.e., the learnable parameters of EA-GNN) and \(\bm{\lambda}\), \textit{alternatively} by minimizing (\ref{loss_func}) and checking the constraint violation. Let \(\bm{\Theta} = \{\Theta_{\text{edge}}, \Theta_{\text{aggr}}, \Theta_{\text{Cheby}}, \Theta_{\text{flat}}\}\) be the collection of the learnable parameters of EA-GNN. Then we have
\begin{equation} \label{theta_update}
    \bm{\Theta}^{\prime} = \operatorname*{argmin}_{\bm{\Theta}}\mathcal{L}_{\text{loss}}. 
\end{equation}
\begin{equation} \label{lambda_update}
    \bm{\lambda} = \Big( \lambda_{c} + \rho\max(0,\nu_{c}(\hat{\mathbf{y}}^{\prime})), \quad c \in \mathcal{C}_{\text{ineq}}\Big), 
\end{equation}

\noindent where \(\hat{\mathbf{y}}^{\prime}\) indicates the output of the hard constraint embedded layer with EA-GNN parameterized by \(\bm{\Theta}^{\prime}\). In other words, it is important to compute (\ref{lambda_update}) using \(\hat{\mathbf{y}}^{\prime}\) which is based on a hot-started \(\bm{\Theta}^{\prime}\) in (\ref{theta_update}) \cite{ref_dnnopf_lag}.

\section{Experimental Settings}\label{section:experimental_settings}
\subsection{Data Preparation}
We use four cases from PGLib (version 23.07) \cite{ref_pglib}: GOC-2312, GOC-3970, GOC-4601, and EPIGRIDS-5658 bus systems, and the real Korean power system (denoted as Korea-4492), provided by Korea Power Exchange (KPX). For data generation, the load data is sampled from a uniform distribution within \([80\%, 120\%]\) of the default value on each bus in GOC-2312, GOC-3970, GOC-4601 and EPIGRIDS-5658. For Korea-4492, we set \([80\%, 110\%]\) considering the electric load pattern of the Korean power systems. Then we sample 10,000 training data and 2,000 test data for each case by using MATPOWER~\cite{ref_matpower}. 

\subsection{Comparative Models}
We extensively compare the proposed GraphOPF with 6 state-of-the-art algorithms, which are summarized as follows.
\begin{itemize}
    \item{PowerModels.jl~\cite{ref_powermodels}: PowerModels.jl is an open-source library in Julia for steady-state power system optimization. It is implemented based on the JuMP~\cite{ref_jump} package, which is a modeling language for mathematical optimization. For solving AC-OPF problems, we use the sparse linear solver MA57 in IPOPT (denoted as IPOPT-MA57 hereafter).}

    \item{ExaModels+MadNLP~\cite{ref_examodels_madnlp}: ExaModels is an algebraic modeling tool for mathematical optimization, specialized for single instruction multiple data (SIMD) abstraction of nonlinear programs. Thus, it supports parallel processing on GPU accelerators. MadNLP is a nonlinear programming solver based on IPOPT. It provides not only various types of sparse and dense linear solvers but also GPU-based solvers (e.g., cuDSS). In our experiments, we use ExaModels and MadNLP in terms GPU.}

    \item{OPF-DNN~\cite{ref_dnnopf_lag}: OPF-DNN follows supervised learning based on a DNN architecture. It outputs \(\mathbf{P}_{g},\mathbf{Q}_{g},\mathbf{V}\), and \(\bm{\theta}\) for given active and reactive load data by minimizing the Lagrangian dual-based loss function.}

    \item{DC3~\cite{ref_dc3}: DC3 is a DL-based nonconvex optimization solver that follows unsupervised learning. In AC-OPF, it first outputs partial variables \(\mathbf{P}_{g}, \mathbf{V}_{g}, \mathbf{V}_{\text{slack}}\) from a DNN model having active/reactive load data as input, and constructs full variables \(\mathbf{P}_{g},\mathbf{Q}_{g},\mathbf{V}, \hspace{1mm}\bm{\theta}\) by using the implicit layer. Note that the AC-OPF problem formulated in \cite{ref_dc3} did not consider the line flow limit, i.e., (\ref{eq:branch_flow_limit}).}

    \item{DeepLDE~\cite{ref_deeplde}: DeepLDE is an unsupervised learning-based AC-OPF solver based on a MLP architecture. Same as DC3, it leverages the implicit layer to construct full variables from partial variables predicted by MLP. To satisfy the inequality constraints, DeepLDE leverages a Lagrangian dual-based loss function.}

    \item{PG-GNN~\cite{ref_gnn_opf6}: PG-GNN is an unsupervised GNN-based AC-OPF solver that leverages the Lagrangian dual-based loss function \cite{ref_dnnopf_lag}. It outputs \(\mathbf{P}_{g},\mathbf{Q}_{g},\mathbf{V}\), and \(\bm{\theta}\) for given input features \(\mathbf{P}_{d}, \mathbf{Q}_{d}, \mathbf{P}_{g}^{\min}, \mathbf{P}_{g}^{\max}. \mathbf{Q}_{g}^{\min}, \mathbf{Q}_{g}^{\max}\). In our experiments, we leverage ChebNet-4 model in \cite{ref_gnn_opf6} that consists of 3 ChebyConv layers with 4 Chebyshev filters.}
\end{itemize}

\subsection{Performance Metrics} 
To evaluate the performance of each model, we use four metrics as follows.
\begin{itemize}
    \item{\textbf{Optimality gap \(\mathcal{F}_{opt}\)}: The optimality gap is computed as \(100 \times {f(\hat{\mathbf{P}}_{g})-f(\mathbf{P}_{g}^{\star})\over{}f(\mathbf{P}_{g}^{\star})}\)} where \(f(\cdot)\) is the generation cost function and \(\hat{\mathbf{P}}_{g}\) is the predicted active power generation of generators. \(\mathbf{P}_{g}^{\star}\) is the cost calculated by the MIPS solver in MATPOWER, which is not necessarily globally optimal but serves as the reference value for performance evaluation.

    % \(\mathcal{Q}_{vio}\)
    \item{\textbf{Feasibility violation}: To check the feasibility of the predicted solution, we compute the feasibility violation (FV) of the inequality constraints (denoted as \(\mathbf{P}_{g}^{FV}, \mathbf{Q}_{g}^{FV}, \mathbf{V}^{FV}\), and \(\mathbf{S}^{FV}\)) and the equality constraints (denoted as \(\text{Active PF}^{FV}\) and \(\text{Reactive PF}^{FV}\)). For inequality constraints, FV is calculated through (\ref{vio_degree_pg_low})$-$(\ref{vio_degree_s}). For equality constraints, FV is the residual between the left and right terms in (\ref{eq:active_power_flow}) and (\ref{eq:reactive_power_flow}), respectively.}
    
    \item{\textbf{Feasibility satisfaction}: We also check the ratios of buses, generators, and transmission lines that satisfy the corresponding inequality (denoted as \(\mathbf{P}_{g}^{FS}, \mathbf{Q}_{g}^{FS}, \mathbf{V}^{FS}\), and \(\mathbf{S}^{FS}\)) and equality (denoted as \(\text{Active PF}^{FS}\) and \(\text{Reactive PF}^{FS}\)) constraints. We set the tolerance as \(10^{-2}\) to the \(\text{Active PF}^{FV}\) and \(\text{Reactive PF}^{FV}\) when calculating the \(\text{Active PF}^{FS}\) and \(\text{Reactive PF}^{FS}\), respectively.}
    
    \item{\textbf{Computational time}: In the case of DL-based methods, NN training time \(T_{train}\) and the inference time \(T_{inf}\) are measured. In the case of mathematical models, the time to get a solution is measured as \(T_{opt}\).}
\end{itemize}

\begin{table}[t]
\caption{Performance comparison with mathematical optimization models for large-scale PGLib power systems.}
\label{math_opt_results}
\begin{adjustbox}{width=\linewidth}
% \begin{center}
% \begin{small}
\begin{sc}
\begin{tabular}{c|c|c|c|c}
\hline
Bus System                  & \begin{tabular}[c]{@{}c@{}}Performance\\ metrics\end{tabular} & \begin{tabular}[c]{@{}c@{}}PowerModels.jl\\ (IPOPT-MA57)\end{tabular} & \begin{tabular}[c]{@{}c@{}}ExaModels+MadNLP \\ (cuDSS)\end{tabular}  & \textbf{GraphOPF} \\ \hline
\multirow{2}{*}{GOC-2312} & \(T_{inf}/T_{opt}\) (\(ms\))     & 4493.90                                                                     & 577                    & \textbf{68.07}                                             \\  
                            & Cost         & \textbf{452259.44}
                                                                & 452264.19                         & 463081.55
                                                                 \\ \hline
\multirow{2}{*}{GOC-3970} & \(T_{inf}/T_{opt}\) (\(ms\))  & 10827.36                                                                       & 1022.72              & \textbf{168.94}                                                           \\  
                            & Cost       & \textbf{3830521.07}                                                                                                                                 & 3830537.23        
                            & 3868293                                                           \\ \hline

\multirow{2}{*}{GOC-4601} & \(T_{inf}/T_{opt}\) (\(ms\))  & 11467.28                                                                         & 1193.87              & \textbf{179.27}                                                    \\  
                            & Cost        & \textbf{3435540.55}                                                                                  & 3435597.09           
                            & 3462860.50                                                             \\ \hline

\multirow{2}{*}{EPIGRIDS-5658} & \(T_{inf}/T_{opt}\) (\(ms\))   & 11279.51                                            & 830.76                 & \textbf{249.73}                                                      \\  
                            & Cost      & \textbf{10150409.46}                                                                         & 10150725.90                   
                                                                & 10271291.50
                                                                 \\ \hline
\end{tabular}
\end{sc}
% \end{small}
% \end{center}
\end{adjustbox}  
\end{table}

\section{Experimental Results}\label{section:experimental_results}
In this section, we provide experimental results for two types of problems: large-scale PGLib power systems and the case study of real Korean power system. Our simulation setup is based on five NVIDIA GPUs (i.e., four NVIDIA RTX 4090 TI GPUs and the NVIDIA A100 GPU) and a Windows desktop with the AMD Ryzen 9 5900X 12-Core Processor and 16 GB RAM. We use five NVIDIA GPUs for NN training, and only use the NVIDIA RTX 4090 TI GPU for the NN test phase and the GPU-based mathematical optimization solver. In the following tables, the values with bold font show the best performance, and red font for unacceptably poor performance, if any.

\begin{table*}[t]
\centering
% \begin{adjustbox}{width=\textwidth}
\caption{Computational time, Optimality gap, and Feasibility satisfaction comparison for large-scale PGLib power systems (DL-based optimization models).}
\label{dl_opt_results}
\begin{adjustbox}{width=12cm}
\begin{sc}
% \begin{center}
\begin{tabular}{c|c|c|c|c|c|c}
\hline
Bus System                  & \begin{tabular}[c]{@{}c@{}}Performance \\ metrics\end{tabular} & PG-GNN & DeepLDE & DC3 & OPF-DNN & \textbf{GraphOPF} \\ \hline
\multirow{8}{*}{GOC-2312} & \(T_{train}\)(\(sec\))/\(N_{train}\)                                                  & 2593/8000           & 7466/1000       & 13021/1000   & 7346/7800       & \textbf{962}/100  \\  
                            & \(\mathcal{F}_{opt}\)(\%)                                                        & \textcolor{red}{26.56}           & 2.86       & \textbf{2.35}   & 24.26       & 2.39    \\  
                            & \(\mathbf{P}_{g}^{FS}\)(\%)                                                        & \textbf{100}        & \textbf{100}       & \textbf{100}   & \textbf{100}       & \textbf{100}   \\  
                            & \(\mathbf{Q}_{g}^{FS}\)(\%)                                                        & \textbf{100}           & 98.78       & 98.15   & \textbf{100}       & 98.88 \\  
                            & \(\mathbf{V}^{FS}\)(\%)                                                         & \textbf{100}           & \textbf{100}       & \textbf{100}   & \textbf{100}       & \textbf{100} \\  
                            & \(\mathbf{S}^{FS}\)(\%)                                                         & \textbf{100}           & 99.89       & 96.11   & 98.89       & 99.97   \\  
                            & \(\text{Active PF}^{FS}\)(\%)                                                 & {42.50}           & \textbf{100}       & \textbf{100}   & \textcolor{red}{39.16}       & \textbf{100}     \\  
                            & \(\text{Reactive PF}^{FS}\)(\%)                                                & \textcolor{red}{27.28}           & \textbf{100}       & \textbf{100}   & 71.05       & \textbf{100}    \\ \hline
\multirow{8}{*}{GOC-3970} & \(T_{train}\)(\(sec\))/\(N_{train}\)                                                    & {5904/8000}           & 17100/1000       &  24660/1000   & 17040/7800               & \textbf{2280}/100 \\  
                            & \(\mathcal{F}_{opt}\)(\%)                                                     & \textcolor{red}{5.91}           & 4.84       & \textbf{0.17}   & 1.97       & 0.99   \\  
                            & \(\mathbf{P}_{g}^{FS}\)(\%)                                                          & {\textbf{100}}           & 99.79       & 99.71   & \textbf{99.99}      & 99.46   \\  
                            & \(\mathbf{Q}_{g}^{FS}\)(\%)                                                         & {\textbf{100}}           & \textcolor{red}{88.32}       & 91.06   & \textbf{99.99}       & 97.48   \\  
                            & \(\mathbf{V}^{FS}\)(\%)                                                          & {\textbf{100}}           & \textbf{100}       & 98.68   & \textbf{100}       & \textbf{100}   \\  
                            & \(\mathbf{S}^{FS}\)(\%)                                                          & {\textbf{100}}           & \textbf{100}       & 99.94   & 99.99       & \textbf{100}    \\  
                            & \(\text{Active PF}^{FS}\)(\%)                                                 & \textcolor{red}{28.02}           & \textbf{100}       & \textbf{100}   & 51.69       & \textbf{100}    \\  
                            & \(\text{Reactive PF}^{FS}\)(\%)                                                & \textcolor{red}{59.27}           & \textbf{100}       & 99.90   & 91.63       & \textbf{100}      \\ \hline
\multirow{8}{*}{GOC-4601} & \(T_{train}\)(\(sec\))/\(N_{train}\)                                                   & {6656/8000}           & 24549/1000       & 25320/1000   & 18060/7800       & \textbf{2271}/100    \\  
                            & \(\mathcal{F}_{opt}\)(\%)                                                    & {4.41}           & \textcolor{red}{7.93}       & \textbf{0.21}   & 5.76       & 0.80      \\  
                            & \(\mathbf{P}_{g}^{FS}\)(\%)                                                & {\textbf{100}}           & 99.84       & \textbf{100}   & \textbf{100}       & \textbf{100}    \\  
                            & \(\mathbf{Q}_{g}^{FS}\)(\%)                                                  & {\textbf{100}}           & \textcolor{red}{86.25}       & 87.22   & \textbf{100}       & 98.29    \\  
                            & \(\mathbf{V}^{FS}\)(\%)                                                 & {\textbf{100}}           & \textbf{\textbf{100}}       & 99.19   & \textbf{100}       & 99.95     \\  
                            & \(\mathbf{S}^{FS}\)(\%)                                                 & {\textbf{100}}           & \textbf{100}       & 99.97   & \textbf{100}       & \textbf{100}    \\  
                            & \(\text{Active PF}^{FS}\)(\%)                                            & \textcolor{red}{24.17}            & \textbf{100}       & 99.96   & 49.88       & \textbf{100}    \\  
                            & \(\text{Reactive PF}^{FS}\)(\%)                                          & \textcolor{red}{62.24}                & \textbf{100}       & 99.96   & 89.69       & \textbf{100}    \\ \hline
\multirow{8}{*}{EPIGRIDS-5658} & \(T_{train}\)(\(sec\))/\(N_{train}\)                                                 & {15209/8000}           & {18372/1000}       & {7451/1000}   & 18809/7800       & \textbf{1991}/100  \\  
                            & \(\mathcal{F}_{opt}\)(\%)                                                  & \textcolor{red}{47.93}           & \textbf{1.01}       & {30.62}   & 17.58       & 1.19     \\  
                            & \(\mathbf{P}_{g}^{FS}\)(\%)                                                   & {\textbf{100}}           & {99.84}       & 99.05   & \textbf{100}       & 99.82   \\  
                            & \(\mathbf{Q}_{g}^{FS}\)(\%)                                                    & {\textbf{100}}           & {76.04}       & \textcolor{red}{53.02}   & \textbf{100}       & 96.13    \\  
                            & \(\mathbf{V}^{FS}\)(\%)                                                       & {\textbf{100}}           & {99.64}       & \textbf{100}   & \textbf{100}       & \textbf{100}     \\  
                            & \(\mathbf{S}^{FS}\)(\%)                                                       & \textcolor{red}{52.80}           & {99.90}       & {99.80}   & 98.31       & \textbf{100}     \\  
                            & \(\text{Active PF}^{FS}\)(\%)                                                  & \textcolor{red}{5.94}           & {\textbf{100}}       & {99.94}   & 32.49       & \textbf{100}      \\  
                            & \(\text{Reactive PF}^{FS}\)(\%)                                                & \textcolor{red}{9.66}           & {\textbf{100}}       & {99.91}   & 49.39     & \textbf{100}       \\ \hline
\end{tabular}
% \end{center}    
\end{sc}
\end{adjustbox}
\end{table*}

\subsection{Performance Comparison Results for Large-scale PGLib Power Systems}\label{section:large-scale ac-opf results}
\subsubsection{Mathematical Optimization Models} \label{section:large-scale ac-opf results_Math}
We compare GraphOPF with PowerModels.jl and ExaModels+MadNLP in terms of average computational time and objective cost. As shown in Table \ref{math_opt_results}, GraphOPF shows up to \(66.02\times\) speedup in computation and \(0.80\%-2.39\%\) of optimality gap for all PGLib power systems. This indicates that GraphOPF can compute substantially faster than the baselines while maintaining the low optimality gap in large-scale AC-OPF. One might think that ExaModels+MadNLP \cite{ref_examodels_madnlp} shows reasonably fast computation and also trustworthy solutions in large-scale AC-OPF, which questions the need of AI-based AC-OPF solutions. However, mathematical optimization models often fail in finding a solution, which will be discussed in Section~\ref{real_acopf_topology_changes} in detail.

\begin{table}[t]
\caption{Performance comparison with mathematical optimization models for Korea-4492 system.}
\label{korea_math_opt_results}
\begin{adjustbox}{width=\linewidth}
% \begin{center}
% \begin{small}
\begin{sc}
\begin{tabular}{c|c|c|c|c}
\hline
Bus System                  & \begin{tabular}[c]{@{}c@{}}Performance\\ metrics\end{tabular} & \begin{tabular}[c]{@{}c@{}}PowerModels.jl\\ (IPOPT-MA57)\end{tabular} & \begin{tabular}[c]{@{}c@{}}ExaModels+MadNLP \\ (cuDSS)\end{tabular}  & \textbf{GraphOPF} \\ \hline
\multirow{2}{*}{Korea-4492} & \(T_{inf}/T_{opt}\) (\(ms\))     & 5630.70                                                                     & 591.14                    & \textbf{259.16}                                             \\  
                            & Cost         & \textbf{121279.59}
                                                                & 121279.71                         & 121317.21
                                                                 \\ \hline
\end{tabular}
\end{sc}
% \end{small}
% \end{center}
\end{adjustbox}  
\end{table}

\subsubsection{DL-based Optimization Models} \label{section:large-scale ac-opf results_DL}
In Table \ref{dl_opt_results}, we compare GraphOPF with PG-GNN \cite{ref_gnn_opf6}, DeepLDE \cite{ref_deeplde}, DC3 \cite{ref_dc3}, and OPF-DNN \cite{ref_dnnopf_lag} in terms of average NN training time, optimality gap, and feasibility satisfaction (FS). Note that distributed data parallelism is used for NN training in the EPIGRIDS-5658 bus system with four NVIDIA RTX 4090 TI GPUs. For all the PGLib power systems, GraphOPF guarantees around 99\% FS on average for the inequality and equality constraints of the generators, buses, and transmission lines. In addition, GraphOPF does not have unacceptably poor performance cases (used \textcolor{red}{red} letters in Table \ref{dl_opt_results}) while other baselines have quite a few cases; for example, PG-GNN exhibits eleven poor cases. This is because the baselines are only verified on small or medium-scale systems, indicating vulnerability to large-scale systems. More importantly, GraphOPF achieves high sample efficiency, leading to fast NN training. As shown in Table~\ref{dl_opt_results}, DeepLDE achieves \(1.04\times\) speedup in NN training time on average for all test cases compared to DC3, indicating the effectiveness of the advanced implicit layer over vanilla one. Notably, GraphOPF can further accelerate this process in \(7.5\!\times-\hspace{1mm}10.81\times\) speedup by reducing the number of training data (denoted as \(N_{train}\)) at least ten times less than the baselines through EA-GNN and the advanced implicit layer.

\subsection{Case Study of Real Power System: Korea Power Exchange} \label{real_time_acopf_topology}
\subsubsection{AC-OPF in a Single Time Instance}
To verify the performance of GraphOPF more practically, we test GraphOPF on the real Korean power system (Korea-4492) with the help of KPX (the only ISO in Korea) in the same way in Section~\ref{section:large-scale ac-opf results}. In Table~\ref{korea_math_opt_results}, GraphOPF shows up to \(21.73\times\) speedup in computation and \(0.03\%\) of optimality gap, indicating that GraphOPF especially performs low optimality gap in real-world data. In addition, as shown in Table~\ref{korea_dl_opt_results}, GraphOPF outperforms DL-based optimization models in terms of NN training time, optimality gap, and FS. Surprisingly, GraphOPF only takes around five minutes for NN training in the Korea-4492 bus system. Thus, this clearly demonstrates that GraphOPF ensures high feasibility, fast training and near-optimal performance even in the real power system, which underpins the practicality of GraphOPF.

\begin{table*}[t]
\centering
% \begin{adjustbox}{width=\textwidth}
\caption{Computational time, Optimality gap, and Feasibility satisfaction comparison for Korea-4492 system (DL-based optimization models).}
\label{korea_dl_opt_results}
\begin{adjustbox}{width=12cm}
\begin{sc}
% \begin{center}
\begin{tabular}{c|c|c|c|c|c|c}
\hline
Bus System                  & \begin{tabular}[c]{@{}c@{}}Performance \\ metrics\end{tabular} & PG-GNN & DeepLDE & DC3 & OPF-DNN & \textbf{GraphOPF} \\ \hline
\multirow{8}{*}{Korea-4492} & \(T_{train}\)(\(sec\))/\(N_{train}\)                                                  & 12370/8000           & 14715/1000       & 78713/1000   & 24977/7800       & \textbf{393}/100  \\  
                            & \(\mathcal{F}_{opt}\)(\%)                                                        & \textcolor{red}{7.37}           & 0.32       & 4.54   & 2.06       & \textbf{0.03}    \\  
                            & \(\mathbf{P}_{g}^{FS}\)(\%)                                                        & \textbf{100}        & 99.82      & 99.85   & \textbf{100}       & 99.83   \\  
                            & \(\mathbf{Q}_{g}^{FS}\)(\%)                                                        & \textbf{100}           & \textbf{100}       & \textbf{100}   & \textbf{100}       & \textbf{100} \\  
                            & \(\mathbf{V}^{FS}\)(\%)                                                         & \textbf{100}           & 99.71       & 98.71   & \textbf{100}       & \textbf{100} \\  
                            & \(\mathbf{S}^{FS}\)(\%)                                                         & 99.91           & \textbf{100}       & \textbf{100}   & 99.98       & \textbf{100}   \\  
                            & \(\text{Active PF}^{FS}\)(\%)                                                 & \textcolor{red}{31.56}           & \textbf{99.99}       & \textbf{99.99}   & 48.84       & \textbf{99.99}     \\  
                            & \(\text{Reactive PF}^{FS}\)(\%)                                                & \textcolor{red}{8.50}           & \textbf{99.91}       & \textbf{99.91}   & 78.39       & \textbf{99.91}    \\ \hline
\end{tabular}
% \end{center}    
\end{sc}
\end{adjustbox}
\end{table*}

\begin{table*}[t]
\centering
\caption{Performance results of GraphOPF for AC-OPF with topology changes (Korean power system).}
\label{real_time_acopf_results}
\begin{adjustbox}{width=\textwidth}
\begin{sc}
\begin{tabular}{c|c|c|c|c|c|c|c|c|c|c|c|c}
\hline
\begin{tabular}[c]{@{}c@{}}Bus System\end{tabular}    & \begin{tabular}[c]{@{}c@{}}\(\mathcal{F}_{opt}\)\\ (\%)\end{tabular} & \begin{tabular}[c]{@{}c@{}}\(\mathbf{P}_{g}^{FS}\)\\ (\%)\end{tabular} & \begin{tabular}[c]{@{}c@{}}\(\mathbf{Q}_{g}^{FS}\)\\ (\%)\end{tabular} & \begin{tabular}[c]{@{}c@{}}\(\mathbf{V}^{FS}\)\\ (\%)\end{tabular} & \begin{tabular}[c]{@{}c@{}}\(\mathbf{S}^{FS}\)\\ (\%)\end{tabular} & \begin{tabular}[c]{@{}c@{}}\(\text{Active PF}^{FS}\)\\ (\%)\end{tabular} & \begin{tabular}[c]{@{}c@{}}\(\text{Reactive PF}^{FS}\)\\ (\%)\end{tabular} & \begin{tabular}[c]{@{}c@{}}\(T_{train}\)/\(T_{inf}\)/\(T_{sol}\)\\ (\(sec\))\end{tabular} & \(N_{train}\), \(N_{test}\) & \(|\mathcal{N}_{g}|\) & \(|\mathcal{N}|\)  & \(|\mathcal{E}|\)  \\ \hline
0100\_49364MW & 0.17                                                  & 99.90                                               & 100                                                 & 100                                              & 100                                              & 99.99                                                    & 99.89                                                      & 54/0.24/3.73                                                            & 15,450           & 158    & 4487 & 5970 \\ \hline
0200\_48129MW & 0.19                                                  & 99.66                                               & 99.87                                                 & 100                                              & 100                                              & 99.99                                                    & 99.88                                                      & 54/0.24/3.72                                                            & 15,450           & 144    & 4486 & 5970 \\ \hline
0300\_47372MW & 0.20                                                  & 99.71                                               & 100                                                 & 100                                              & 100                                              & 99.99                                                    & 99.90                                                      & 53/0.24/3.54                                                            & 15,450           & 145    & 4486 & 5970 \\ \hline
0400\_47320MW & 0.21                                                  & 99.80                                               & 100                                                 & 100                                              & 100                                              & 99.99                                                    & 99.90                                                      & 54/0.24/3.60                                                            & 15,450           & 147    & 4486 & 5970 \\ \hline
0500\_47943MW & 0.18                                                  & 99.86                                               & 100                                                 & 100                                              & 100                                              & 99.99                                                    & 99.89                                                      & 53/0.24/3.71                                                            & 15,450           & 153    & 4485 & 5969 \\ \hline
0600\_50000MW & 0.31                                                  & 99.90                                               & 100                                                 & 99.97                                              & 100                                              & 99.99                                                    & 99.90                                                      & 54/0.24/3.74                                                            & 15,450           & 168    & 4486 & 5970 \\ \hline
0700\_53639MW & 0.23                                                  & 99.88                                               & 100                                                 & 99.98                                              & 100                                              & 99.99                                                    & 99.90                                                      & 54/0.24/3.71                                                            & 15,450           & 185    & 4524 & 6030 \\ \hline
\textcolor{red}{0800\_55646MW} & -                                                  & 99.94                                               & 99.96                                                 & 99.98                                              & 100                                              & 99.99                                                    & 99.88                                                      & 102/0.24/$\textcolor{red}{\infty}$                                                            & 15,450           & 203    & 4529 & 6029 \\ \hline
0900\_58364MW & 0.11                                                  & 99.82                                               & 100                                                 & 99.97                                              & 99.99                                              & 99.99                                                    & 99.89                                                      & 105/0.25/4.39                                                             & 15,450           & 212    & 4526 & 6027 \\ \hline
1000\_56322MW & 0.22                                                  & 99.90                                               & 99.98                                                 & 99.96                                              & 99.99                                              & 99.99                                                    & 99.90                                                      & 57/0.25/4.16                                                            & 15,450           & 214    & 4522 & 6018 \\ \hline
1100\_53234MW & 0.25                                                  & 99.94                                               & 99.73                                                 & 100                                              & 100                                              & 99.99                                                    & 99.90                                                      & 103/0.24/4.62                                                            & 15,450           & 193    & 4510 & 6006 \\ \hline
\textcolor{red}{1204\_49120MW} & -                                                  & 99.92                                               & 100                                                 & 99.95                                              & 100                                              & 99.98                                                    & 99.90                                                      & 102/0.24/$\textcolor{red}{\infty}$                                                            & 15,450           & 185    & 4514 & 6009 \\ \hline
\textcolor{red}{1304\_49184MW} & -                                                  & 99.74                                               & 100                                                 & 99.91                                              & 100                                              & 99.99                                                    & 99.90                                                      & 109/0.25/$\textcolor{red}{\infty}$                                                            & 15,450           & 192    & 4515 & 6011 \\ \hline
\textcolor{red}{1404\_50855MW} & -                                                  & 99.96                                               & 100                                                 & 99.91                                              & 100                                              & 99.99                                                    & 99.90                                                      & 54/0.24/$\textcolor{red}{\infty}$                                                           & 15,450           & 202    & 4516 & 6013 \\ \hline
1503\_50930MW & 0.15                                                  & 99.80                                               & 100                                                 & 100                                              & 100                                              & 99.99                                                    & 99.90                                                      & 102/0.24/4.76                                                            & 15,402           & 214    & 4522 & 6016 \\ \hline
1603\_52305MW & 0.84                                                  & 99.86                                               & 99.97                                                 & 99.95                                              & 100                                              & 99.99                                                    & 99.90                                                      & 48/0.21/4.50                                                            & 15,450           & 224    & 4527 & 6028 \\ \hline
1703\_54300MW & 0.84                                                  & 99.79                                               & 99.78                                                 & 99.91                                              & 100                                              & 99.99                                                    & 99.90                                                      & 56/0.23/4.51                                                            & 15,450           & 226    & 4537 & 6038 \\ \hline
1803\_56622MW & 0.24                                                  & 99.90                                               & 99.98                                                 & 100                                              & 100                                              & 99.99                                                    & 99.90                                                      & 48/0.21/4.25                                                            & 15,450           & 219    & 4536 & 6040 \\ \hline
1903\_58834MW & 0.18                                                  & 99.97                                               & 100                                                 & 99.98                                              & 100                                              & 99.99                                                    & 99.90                                                      & 49/0.21/4.32                                                            & 15,448           & 218    & 4538 & 6044 \\ \hline
2004\_58790MW & 0.32                                                  & 99.81                                               & 100                                                 & 100                                              & 100                                              & 99.99                                                    & 99.90                                                      & 91/0.21/4.26                                                            & 15,442           & 213    & 4538 & 6046 \\ \hline
\textcolor{red}{2104\_57559MW} & -                                                  & 99.95                                               & 99.97                                                 & 99.95                                              & 100                                              & 99.99                                                    & 99.91                                                      & 48/0.21/$\textcolor{red}{\infty}$                                                            & 15,450           & 203    & 4505 & 5994 \\ \hline
2203\_56467MW & 0.51                                                  & 99.75                                               & 99.98                                                 & 100                                              & 100                                              & 99.99                                                    & 99.91                                                      & 49/0.21/4.38                                                            & 15,450           & 199    & 4503 & 5992 \\ \hline
2304\_55651MW & 0.29                                                  & 99.80                                               & 100                                                 & 100                                              & 100                                              & 99.99                                                    & 99.88                                                      & 103/0.23/4.39                                                            & 15,450           & 184    & 4496 & 5986 \\ \hline
\end{tabular}
\end{sc}
\end{adjustbox}
\end{table*}

\subsubsection{AC-OPF with Topology Changes} \label{real_acopf_topology_changes}
The most challenging part of AI AC-OPF comes from the topology-fragile nature. To verify the adaptability against topology change, we test GraphOPF on the Korean power system for entire one day of historical operation. Table~\ref{real_time_acopf_results} summarizes the performance of GraphOPF in terms of average NN training time, optimality gap, and FS. As can be seen, the numbers of generators (denoted as \(|\mathcal{N}_{g}|\)) also changes ranging from 144 to 226. Similarly, the number of buses and transmission lines change. 

Considering GNN architecture can solve AC-OPF with topology changes~\cite{ref_gnn_opf1, ref_gnn_opf2, ref_gnn_opf4, ref_gnn_opf6}, we leverage transfer learning with frozen layers that only transfer the learnable parameters of EA-GNN layers (i.e., fine-tuning the flatten layer). In doing this, we find that NN training of GraphOPF can be further accelerated and thus finished less than one minute; surprisingly, only 15 training samples were enough for fine-tuning in the case of Korean power system. We also find that GraphOPF achieves 99\% FS on average for the five cases where the mathematical solvers fail to converge, and thus the solving time denoted by \(T_{sol}\) is recorded as \(\infty\) (\textcolor{red}{red} letters in Table~\ref{real_time_acopf_results}). In addition, the cost gap (denoted by \(\mathcal{F}_{opt}\)) between the GraphOPF and the MATPOWER solver cannot be recorded for those five cases. This shows that GraphOPF can be an alternative method for critical situations during power system operation, which may happen frequently in future power systems due to the high penetration of DERs and load growth.

\section{Ablation Studies}
GraphOPF consists of EA-GNN and a hard constraint embedded layer. Especially, EA-GNN is designed with MEConv and ChebyConv layers, cooperatively capturing generalized topological features even with a small amount of data on large-scale power systems by additionally using edge features. To support this, ablation studies are conducted by eliminating an MEConv layer of EA-GNN (called M1 hereafter).

\begin{figure*}[t]
\centering
\includegraphics[width=14.5cm]{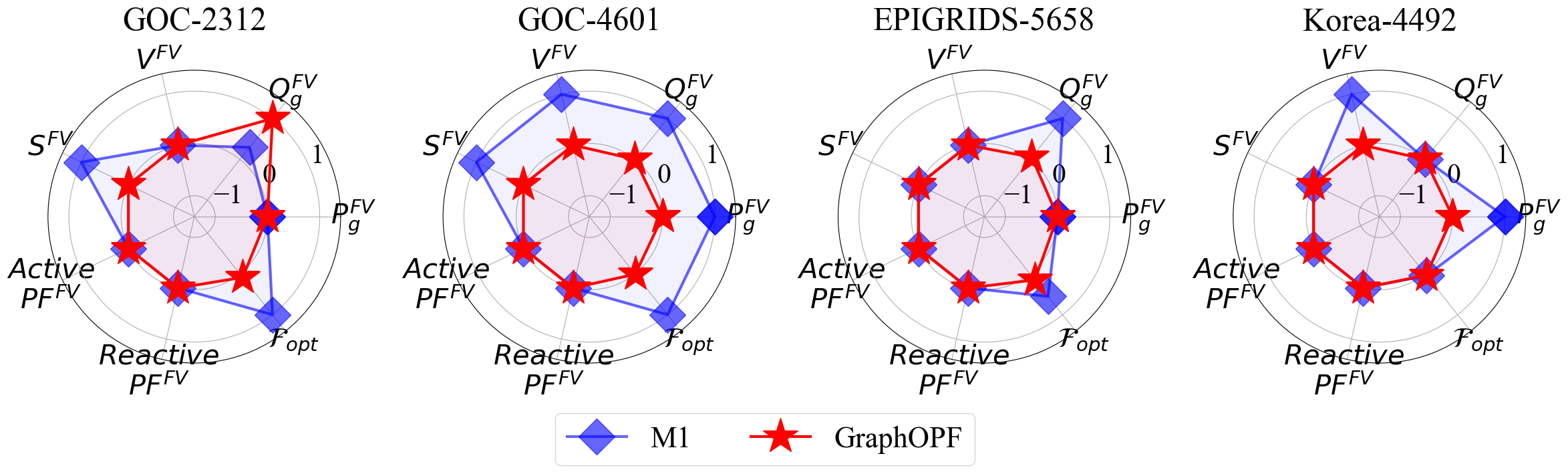}
\caption{Radar charts for optimality gap and feasibility violation results of ablation studies.}
\label{ablation_performance}
\end{figure*}

\subsection{Performance Comparison Results}
\cref{ablation_performance} shows the performance results of optimality gap and FV for GOC-2312, GOC-4601, EPIGRIDS-5658,
and Korea-4492 bus systems. Note that M1 fails in model training for GOC-3970. As can be seen, GraphOPF outputs near-optimal and feasible solutions, showing that MEConv layer is effective for large-scale AC-OPF.

\subsection{Feature Drift Analysis for Generalization Performance}
Table~\ref{feature_drift_table} presents the generalization performance by measuring the feature drift on two operating scenarios of AC-OPF in Korea power system: maximum and minimum load cases in a fixed topology and topology changing cases with different loads. The details of operating scenarios are in Table~\ref{operating_scenarios}. First, the features are extracted by EA-GNN layers from GraphOPF and ChebyConv layers from M1 for each case in the scenario, respectively. Then, feature drift is computed for case 1 and 2 by using graph mean pooling~\cite{ref_gnn_book} and Sinkhorn-Wasserstein distance~\cite{cuturi2013sinkhorn}. For graph mean pooling, we use the ratio of normalized L2 norm to compute feature drift as \(100\times{{2\|\mathbf{F}^{(1)}-\mathbf{F}^{(2)}\|_{2}}\over{\|\mathbf{F}^{(1)}\|_{2}+\|\mathbf{F}^{(2)}\|_{2}}}\), where \(\mathbf{F}^{(1)}\) and \(\mathbf{F}^{(2)}\) are graph-level features for case 1 and 2 of operating scenario. In Table~\ref{feature_drift_table}, EA-GNN shows invariant feature drift, demonstrating robust generalization performance against diverse operating scenarios. Interestingly, increasing the number of training samples to tenfold for M1 (denoted as M1-1000) is helpful to reduce feature drift. Nevertheless, EA-GNN still outperforms M1-1000. Hence, it demonstrates that EA-GNN is enough to achieve generalization by using only 100 training samples, and fine-tuning on flatten layer while freezing EA-GNN layers can be effective on topology changes.

\begin{table}[t]
\centering
\caption{Feature drift results on two operating scenarios.}
\label{feature_drift_table}
\begin{adjustbox}{width=7.5cm}
\begin{sc}
\begin{tabular}{|ccc}
\hline
% \multicolumn{1}{c|}{Model}   & \multicolumn{1}{c|}{\begin{tabular}[c]{@{}c@{}}Normalized L2 norm\\ (\%)\end{tabular}} & Sinkhorn-Wassertein distance \\ \hline
\multicolumn{1}{c|}{Model}   & \multicolumn{1}{c|}{\begin{tabular}[c]{@{}c@{}}Normalized L2 norm\\ (\%)\end{tabular}} & \multicolumn{1}{c}{\begin{tabular}[c]{@{}c@{}}Sinkhorn-Wasserstein \\ distance \end{tabular}} \\ \hline
\multicolumn{3}{c}{Fixed Topology}                                                                                                                  \\ \hline
\multicolumn{1}{c|}{\textbf{EA-GNN}}  & \multicolumn{1}{c|}{\textbf{0.0026}}                                                          & \textbf{0.0023}                     \\ \hline
\multicolumn{1}{c|}{M1}      & \multicolumn{1}{c|}{6.89}                                                              & 7.39                         \\ \hline
\multicolumn{1}{c|}{M1-1000} & \multicolumn{1}{c|}{5.21}                                                              & 5.52                         \\ \hline
\multicolumn{3}{c}{Topology Changes}                                                                                                                \\ \hline
\multicolumn{1}{c|}{\textbf{EA-GNN}}  & \multicolumn{1}{c|}{\textbf{0.065}}                                                             & \textbf{0.0024}                     \\ \hline
\multicolumn{1}{c|}{M1}      & \multicolumn{1}{c|}{13.83}                                                             & 13.48                        \\ \hline
\multicolumn{1}{c|}{M1-1000} & \multicolumn{1}{c|}{16.32}                                                             & 11.80                        \\ \hline
\end{tabular}
\end{sc}
\end{adjustbox}
\end{table}

\begin{table}[t]
\centering
\caption{Two operating scenarios of AC-OPF in Korea power system.}
\label{operating_scenarios}
\begin{adjustbox}{width=5.5cm}
\begin{sc}
\begin{tabular}{|ccccc}
\hline
\multicolumn{1}{c|}{Scenarios} & \multicolumn{1}{c|}{\begin{tabular}[c]{@{}c@{}}Load \\ (MW)\end{tabular}} & \multicolumn{1}{c|}{\(|\mathcal{N}_{g}|\)}                   & \multicolumn{1}{c|}{\(|\mathcal{N}|\)}                     & \(|\mathcal{E}|\)                     \\ \hline
\multicolumn{5}{c}{Fixed Topology}                                                                                                                                                                                         \\ \hline
\multicolumn{1}{c|}{Case1}     & \multicolumn{1}{c|}{51408}                                                & \multicolumn{1}{c|}{\multirow{2}{*}{182}} & \multicolumn{1}{c|}{\multirow{2}{*}{4492}} & \multirow{2}{*}{5969} \\ \cline{1-2}
\multicolumn{1}{c|}{Case2}     & \multicolumn{1}{c|}{50064}                                                & \multicolumn{1}{c|}{}                     & \multicolumn{1}{c|}{}                      &                       \\ \hline
\multicolumn{5}{c}{Topology Changes}                                                                                                                                                                                       \\ \hline
\multicolumn{1}{c|}{Case1}     & \multicolumn{1}{c|}{58834}                                                & \multicolumn{1}{c|}{218}                  & \multicolumn{1}{c|}{4538}                  & 6044                  \\ \hline
\multicolumn{1}{c|}{Case2}     & \multicolumn{1}{c|}{47320}                                                & \multicolumn{1}{c|}{147}                  & \multicolumn{1}{c|}{4486}                  & 5970                  \\ \hline
\end{tabular}
\end{sc}
\end{adjustbox}
\end{table}

\section{Conclusion}\label{section:conclusion}
In this paper, we propose a fast, topology-adaptable, scalable, and unsupervised physics-informed graph learning framework called GraphOPF that considers large-scale AC-OPF. By combining EA-GNN with a hard constraint embedded layer, the NN training time of GraphOPF was substantially accelerated while satisfying the constraint feasibility. Extensive experimental results demonstrated that GraphOPF achieves feasible and near-optimal AC-OPF solutions in large-scale power systems, including the real Korean power system. Furthermore, GraphOPF shows remarkable performance in AC-OPF problems with topology changes by leveraging transfer learning in EA-GNN layers. This improves sample efficiency in GraphOPF to accelerate NN training further, reducing it less than one minute in Korean power system. In our future work, we will consider graph transfer learning in GraphOPF to build an OPF foundation model.

\section*{Acknowledgements}
This work was supported by the National Research Foundation of Korea (NRF) funded by the Ministry of Science and ICT under Grant RS-2025-02215243.

\nocite{langley00}

\bibliography{example_paper}
\bibliographystyle{icml2026}

% %%%%%%%%%%%%%%%%%%%%%%%%%%%%%%%%%%%%%%%%%%%%%%%%%%%%%%%%%%%%%%%%%%%%%%%%%%%%%%%
% %%%%%%%%%%%%%%%%%%%%%%%%%%%%%%%%%%%%%%%%%%%%%%%%%%%%%%%%%%%%%%%%%%%%%%%%%%%%%%%
% % APPENDIX
% %%%%%%%%%%%%%%%%%%%%%%%%%%%%%%%%%%%%%%%%%%%%%%%%%%%%%%%%%%%%%%%%%%%%%%%%%%%%%%%
% %%%%%%%%%%%%%%%%%%%%%%%%%%%%%%%%%%%%%%%%%%%%%%%%%%%%%%%%%%%%%%%%%%%%%%%%%%%%%%%
% \newpage
% \appendix
% \onecolumn

% You can have as much text here as you want. The main body must be at most $8$
% pages long. For the final version, one more page can be added. If you want, you
% can use an appendix like this one.

% The $\mathtt{\backslash onecolumn}$ command above can be kept in place if you
% prefer a one-column appendix, or can be removed if you prefer a two-column
% appendix.  Apart from this possible change, the style (font size, spacing,
% margins, page numbering, etc.) should be kept the same as the main body.
% %%%%%%%%%%%%%%%%%%%%%%%%%%%%%%%%%%%%%%%%%%%%%%%%%%%%%%%%%%%%%%%%%%%%%%%%%%%%%%%
% %%%%%%%%%%%%%%%%%%%%%%%%%%%%%%%%%%%%%%%%%%%%%%%%%%%%%%%%%%%%%%%%%%%%%%%%%%%%%%%

\end{document}